\documentclass[11pt,twoside]{article}

\usepackage{asp2014}

\aspSuppressVolSlug
\resetcounters

\begin{document}

\title{Metadata and their importance in SO/PHI's on-board data processing}

\author{K. Albert,$^1$ J. Hirzberger,$^1$ D. Busse,$^1$ J. S. Castellanos Dur\'{a}n,$^1$ P. Gutierrez-Marques,$^1$ and M. Kolleck,$^1$
\affil{$^1$Max Planck Institute for Solar System Research, G\"{o}ttingen, Germany \email{albert@mps.mpg.de}}}


\newcommand{\helix}[1]{{H\tiny{E}\normalsize{LI}\tiny{X}$^{{\textstyle+}}$#1}} 

\begin{abstract}
To cope with the telemetry limitations, the Polarimetric and Helioseismic Imager on Solar Orbiter does full on-board data processing. Metadata are central to the autonomous processing flow, crucial for providing science ready data sets to the community, as well as important in the blind debugging process that will occur in the commissioning phase. We designed a custom metadata logging system for SO/PHI. This paper shows how the logged information is used in the blind debugging scenario.
\end{abstract}



\section{Introduction}
State of the art scientific instrumentation, especially those deployed in deep space, often produce more data than can be downloaded. This is the case for the Polarimetric and Helioseismic Imager \citep[PHI,][]{Solanki_PHI} on-board the Solar Orbiter \citep[SO,][]{Mueller2013_SO} spacecraft. SO/PHI is an imaging spectropolarimeter, recording four-million-pixel images at six wavelengths in four polarisation states to retrieve five physical quantities: magnetic field strength, inclination, azimuth, line-of-sight velocity and temperature map. The limitations on telemetry from Solar Orbiter would allow downloading $\sim$30 raw science data sets in each orbit. An orbit ($\sim$160 days) typically accommodates 30\,days of observations at strategic points, therefore this would mean little data return. In addition, due to accuracy requirements of SO/PHI, instrument characterisation must be done in orbit, right before the observations, a significant addition to the necessary telemetry.

To maximise science return and cope with telemetry constraints we implemented full on-board data processing in SO/PHI \citep[see][]{Albert2018_Autonomous,lange2017board}. The instrument calculates operational parameters for data acquisition, determines calibration data, which is then applied to the science data sets, before reducing them to the final physical quantities of interest by inverting the radiative transfer equation (RTE, see Fig.\,\ref{pipe_fig}). These steps are done for the first time in orbit, with severely limited hardware when compared to ground processing  \citep[see][]{Albert2018_Performance}, without free entry points for verification and without full access to partial results.

\articlefigure[width=0.6\textwidth]{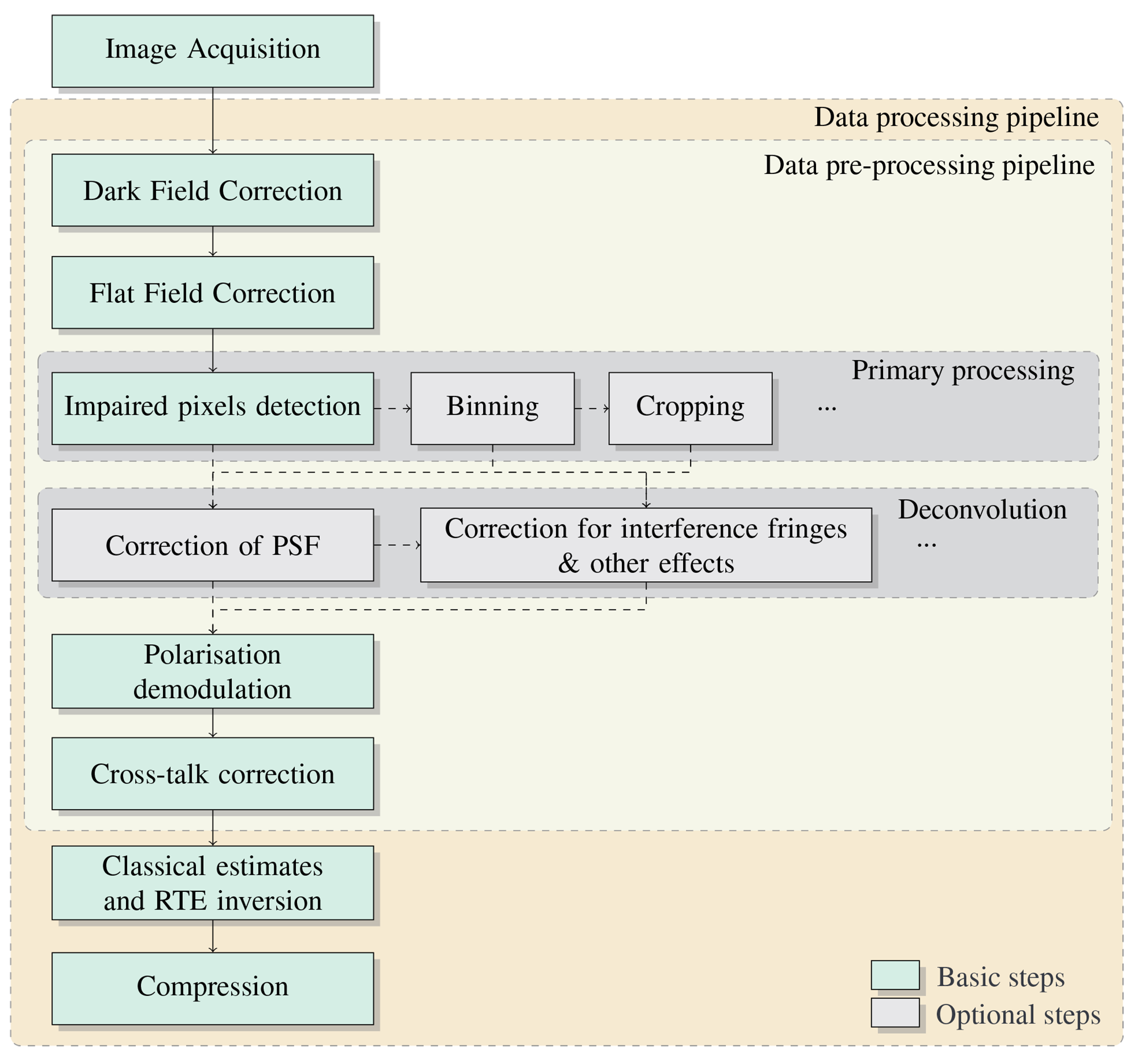}{pipe_fig}{Typical processing pipeline for spectropolarimetric science data with compulsory and optional steps.}

Metadata play a central role in the design of the on-board data processing system. Each data set has its own associated metadata file, created at image acquisition, containing all hardware parameters and processing plans. All processing steps add their own entries to this information, generating a full log of the processing. This file is then used both on-board and on ground.

\section{The metadata system}
Metadata are crucial to the success of SO/PHI. Due to the on-board processing, the most important source of information on the data reduction details is what we record in the process. As the data are for the wide scientific community, the metadata must describe the data sets in their entirety, including all information necessary for their scientific use.

\subsection{Recording}
There are four different sources of metadata: the planning process, the calibration campaign, the instrument, and the data processing system. In the planning process we define processing parameters, such as the data set identifiers for calibration data. We write these into the so called processing environment, which is a file located on-board. The processing environment is further extended during the calibration campaign with other calculated values. It is then written into the metadata of the data set at acquisition, together with the instrument parameters. During the processing we enlist all operations performed with their parameters and return values, followed by higher level information regarding the reason for those steps, as well as a data set summary, showing the current parameters of the data set. Additionally, at steps where we load additional calibration data, the hardware parameters at the recording of the two data sets are also cross-checked to generate warnings for ground review. For pixel-wise information we use an additional image, treated as bit mask, to encode the pixels that reached a \texttt{NaN} value during processing at any point in time, and other areas of interest. Before data download this mask is encoded into the temperature image, where we do not loose unrecoverable information by doing so.

\subsection{On-board use}
The processing pipeline can be executed with the parameters recorded in the metadata of the data sets, or with the current processing environment. Each step of the processing is also based on metadata. We check the basic parameters of the data set: how many images does it contain, which area of the detector is it from, was it binned, and how is it scaled. These values determine further actions, e.g. how will we scale the data set to ensure accuracy in the upcoming operations, or the necessity of cropping or binning calibration data. The fact that this information is carried in the metadata ensures that no additional information must be passed from one step of the pipeline to the next, and the data set can be understood at any time independently from the pipeline. 

\subsection{On-ground use}
After the data download we check whether there are any errors or warnings, and at which step did they occur, if any. At the time of commissioning we may do "blind debugging": find any error that occurred without direct access to the pipeline parameters at runtime. In such a case the metadata will provide us with information necessary to find the error source. During this period partial results will also be available to reproduce any problem encountered in flight on a ground instrument model.

Once the data set is available to the science community, some of the interests are the steps taken during the on-board processing, and the accuracy of the data set (reconstructed from the logs). In addition, pixel-wise information is also available from the masks to ensure that potentially new discoveries are not instrument artefacts. 

\section{Example for blind debugging}
During blind debugging we use the higher level metadata. These contain an entry by each pipeline block, with name, time of execution, operation target, input parameters, data set dimensions, and return value. In addition to this we also have lower level metadata and data set summaries available.

An example for erroneous results is shown in Fig.\,\ref{results_fig}, alongside the expected results for comparison. The metadata associated with the results is shown in Fig.\,\ref{meta_fig}. From this information it is possible to assess that there was a Feed Select Mechanism (FSM) mismatch between the processed image and the demodulation matrix, indicating that the two optical paths are not identical. It is also visible that the OperandID, referring to the ID of the demodulation matrix is not the expected one, hence there was an operator error.

\articlefigure[width=.9\textwidth]{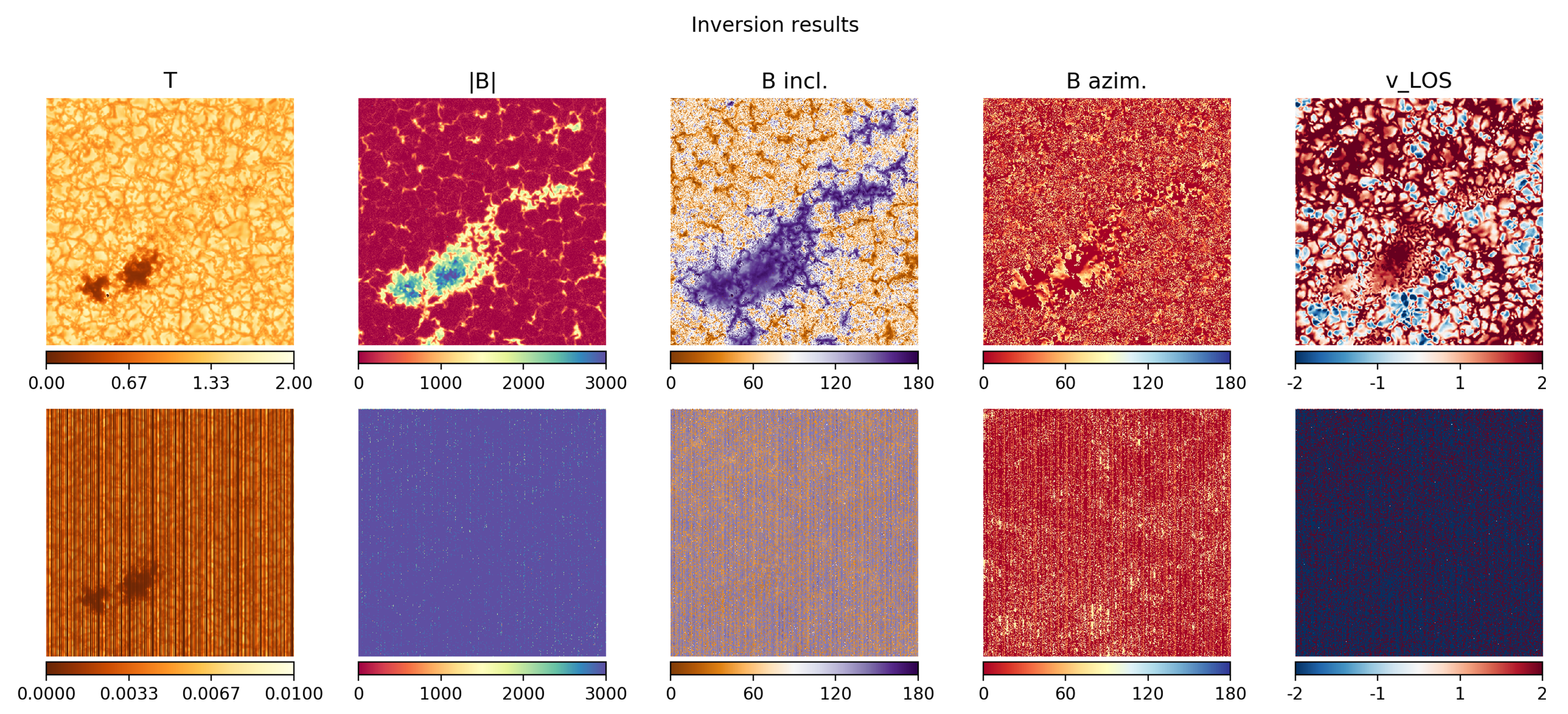}{results_fig}{\textit{Top}: Expected results. \textit{Bottom}: Obtained, erroneous results. Data is from magnetohydrodynamics simulations \citep{Tino2017Inv}, prepared with our instrument simulator \citep{blanco2018sophism}. For testing purposes the RTE is inverted with \helix. \citep{Lagg2004}.}
\articlefigure[width=\textwidth, trim={.3cm 0 0 0},clip]{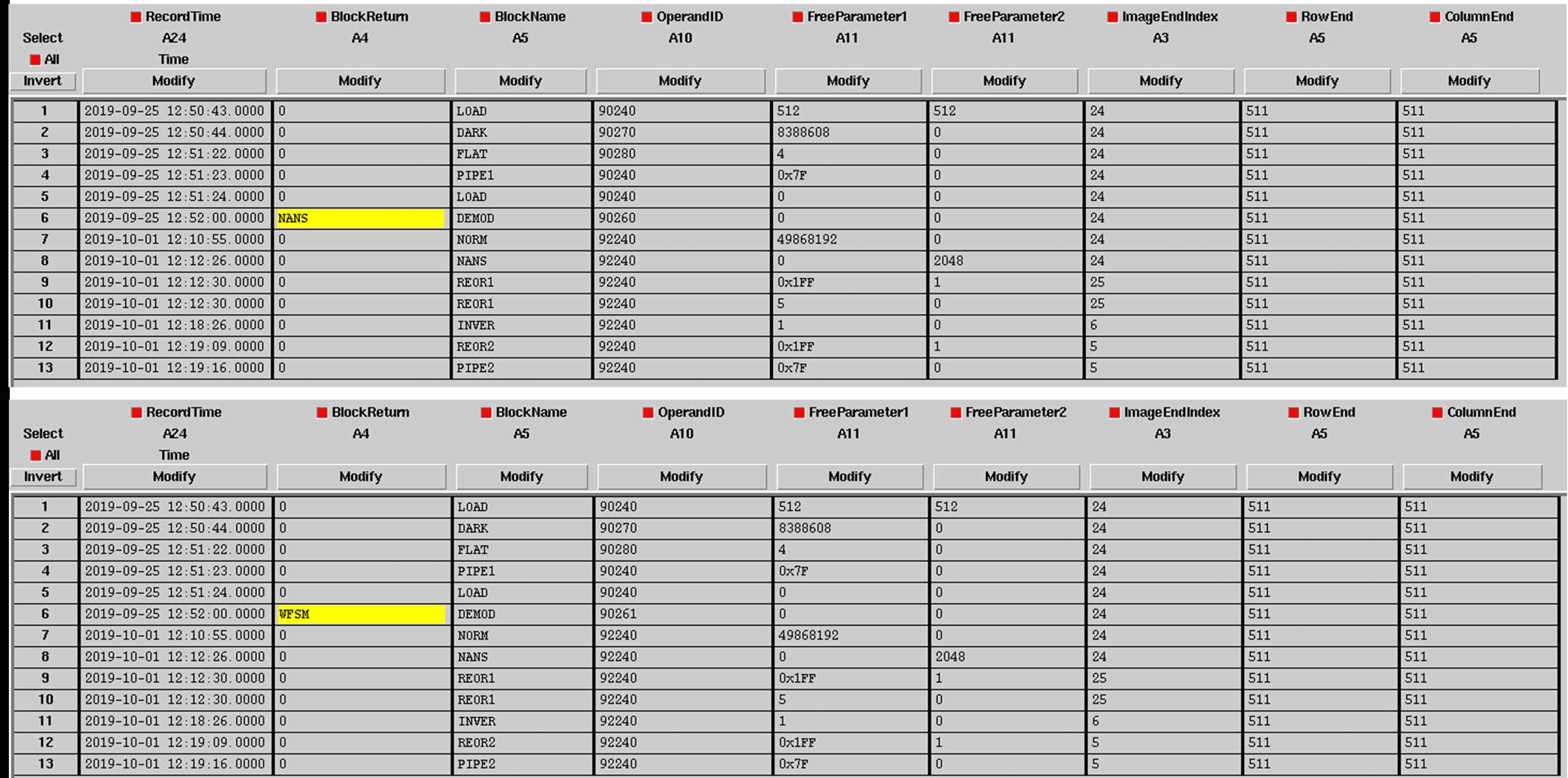}{meta_fig}{Excerpt of recorded metadata. \textit{Top}: From the expected results, indicating that the result contains NaNs, as expected. \textit{Bottom}: From erroneous results, with warning regarding the Feed Select Mechanism, and incorrect OperandID.}

\section{Conclusions}
Metadata are central to the success of SO/PHI. We have custom designed our metadata collection system, making it not only contain all essential information for ground use, but also be the central source of information for the processing pipelines. An example of ground use in blind debugging is presented, to illustrate how the recorded data give clues on what could have gone wrong during processing.

\acknowledgements Workframe: International Max Planck Research School (IMPRS) for Solar System Science. Solar Orbiter: ESA, NASA. Grant: DLR 50 OT 1201.

\bibliography{P10-26}  

\begin{thebibliography}{}
\expandafter\ifx\csname natexlab\endcsname\relax\def\natexlab#1{#1}\fi
\expandafter\ifx\csname url\endcsname\relax
  \def\url#1{\texttt{#1}}\fi
\expandafter\ifx\csname urlprefix\endcsname\relax\def\urlprefix{URL }\fi
\providecommand{\eprint}[2][]{\url{#2}}

\bibitem[{Albert et~al.(2018{\natexlab{a}})Albert, Hirzberger, Busse
  et~al.}]{Albert2018_Autonomous}
Albert, K., Hirzberger, J., Busse, D., et~al. 2018{\natexlab{a}}, in Proc.
  SPIE, vol. 707, 10707

\bibitem[{Albert et~al.(2018{\natexlab{b}})Albert, Hirzberger, Busse
  et~al.}]{Albert2018_Performance}
--- 2018{\natexlab{b}}, in ASP Conference Series, Vol. 523

\bibitem[{{Blanco Rodr{\'\i}guez} et~al.(2018){Blanco Rodr{\'\i}guez}, {del
  Toro Iniesta}, \& {Orozco Su{\'a}rez}}]{blanco2018sophism}
{Blanco Rodr{\'\i}guez}, J., {del Toro Iniesta}, J.~C., \& {Orozco Su{\'a}rez},
  D. e.~a. 2018, \apjs, 237, 35

\bibitem[{{Lagg} et~al.(2004){Lagg}, {Woch}, {Krupp}, \& {Solanki}}]{Lagg2004}
{Lagg}, A., {Woch}, J., {Krupp}, N., \& {Solanki}, S.~K. 2004, Astronomy and
  Astrophysics, 414, 1109

\bibitem[{Lange et~al.(2017)Lange, Fiethe, Michel et~al.}]{lange2017board}
Lange, T., Fiethe, B., Michel, H., et~al. 2017, in {NASA/ESA} Conf. on Adapt.
  Hardw. and Sys.

\bibitem[{M\"{u}ller et~al.(2013)M\"{u}ller, Marsden, Cyr
  et~al.}]{Mueller2013_SO}
M\"{u}ller, D., Marsden, R.~G., Cyr, O.~S., et~al. 2013, Solar Physics, 285, 25

\bibitem[{Riethm\"{u}ller et~al.(2017)Riethm\"{u}ller, Solanki, Barthol
  et~al.}]{Tino2017Inv}
Riethm\"{u}ller, T.~L., Solanki, S.~K., Barthol, P., et~al. 2017, The
  Astrophysical Jour. Suppl. Series

\bibitem[{Solanki et~al.(2018)Solanki, del Toro~Iniesta, Woch
  et~al.}]{Solanki_PHI}
Solanki, S., del Toro~Iniesta, J., Woch, J., et~al. 2018, Accepted to Astronomy
  and Astrophysics

\end{thebibliography}


\end{document}